\documentstyle[eqsecnum,aps,epsf,twocolumn]{revtex}
\begin{document}
\title{Multicanonical Chain Growth Algorithm}
\author{Michael Bachmann\thanks{Email address: michael.bachmann@itp.uni-leipzig.de} 
and Wolfhard Janke\thanks{Email address: wolfhard.janke@itp.uni-leipzig.de}}
\address{Institut f\"ur Theoretische Physik, Universit\"at Leipzig,
Augustusplatz 10/11, D-04109 Leipzig, Germany}
\maketitle
\begin{abstract}
We present a temperature-independent Monte Carlo method for the determination of the
density of states of lattice proteins that combines the fast ground-state search strategy
of the nPERM chain growth and multicanonical reweighting for sampling
the complete energy space. Since the density of states contains all energetic information
of a statistical system, we can directly calculate the mean energy, specific heat, Gibbs free energy, 
and entropy for all temperatures. We apply this method to HP lattice
proteins and for the examples of sequences considered, we identify the transitions 
between native, globule, and random coil states. Since no special properties of
heteropolymers are involved in this algorithm, the method applies to polymer models
as well. 
\end{abstract}
\vspace{8mm}
The simulation of protein folding is extremely challenging, since the interactions between the 
constituents of the macromolecule and the influence of the environment require sophisticated
models. One of the most essential aspects in the description of the folding process is the
formation of a compact core of hydrophobic amino acid residues (H) which is screened from water by
hydrophilic or polar residues (P). This characteristic property of realistic proteins  
can be qualitatively studied with simple lattice models such as the HP model~\cite{dill1}.
By taking into account the attractive interaction
between hydrophobic monomers only, the energy of a lattice protein 
with certain conformation and sequence is calculated as follows:
\begin{equation}
\label{hpmodel}
E = -\sum\limits_{\langle i,j<i-1\rangle} \sigma_i\sigma_j,
\end{equation}
where $\langle i,j<i-1\rangle$ symbolises that the sum is taken only over nearest lattice 
neighbours being nonadjacent along the self-avoiding chain of monomers. If the $i$th monomer is hydrophobic, 
$\sigma_i = 1$, while for a polar monomer $\sigma_i = 0$.

As it is one of the main goals in off-lattice simulations to find low-lying energy states 
within a rough free energy landscape, good lattice folders are expected to have low-degenerated
ground states. Much work has been done on identifying designing sequences with such native 
states. Ground-state search strategies on three-dimensional lattices range, for example, from 
enumeration~\cite{tang1,irbaeck1} over hydrophobic core construction~\cite{dill2,dill3} and
contact interaction~\cite{toma1} to chain growth methods~\cite{dill4,grassberger1,grassberger2,hsu1}. 
Low-lying energy states for HP sequences with up to 136 monomers were identified with these
methods.

In contrast, there were only few attempts to study the thermodynamic properties of the
HP model in three dimensions~\cite{iba1}. 
The main reason is that conventional Monte
Carlo methods like Metropolis sampling, but also more sophisticated methods like
simulated~\cite{simtemp} and parallel~\cite{paratemp} tempering 
as well as histogram reweighting Monte Carlo
algorithms such as multicanonical sampling~\cite{muca} or the Wang-Landau method~\cite{wanglandau}
expose problems in tackling
``hidden'' conformational barriers in combination with chain update moves which become
usually inefficient at low temperatures, where many attempted moves are rejected due to 
the self-avoidance constraint.  One possibility to update the conformation is to apply move sets.
Widely used sets usually consist of operations that change a single bond (end flips), two bonds
(corner flips), three (crankshaft) or even more bonds, and pivot rotations, where the
$i$th monomer serves as pivot point and one of the two partial chains connected with it
is rotated about any axis through the pivot. Unfortunately, the more dense the conformation, the
more inefficient this procedure becomes.

Alternatively, it is possible to let the polymer grow, i.e., the $n$th monomer is placed
at a randomly chosen next-neighbour site of the $(n-1)$th monomer ($n\le N$ with $N$
being the total length of the polymer). If this new site is already occupied,
the entire chain would have to be discarded to obtain correct statistics. This simple chain
growth is not efficient, since the number of discarded chains grows exponentially with the
chain length. Rosenbluth chain growth~\cite{rosenbluth} avoids occupied neighbours at the expense of
a bias, since the probability of such a chain is $p_n\sim \left(\prod_{l=2}^n m_l\right)^{-1}$, where $m_l$ is
the number of free lattice sites to place the $l$th monomer. This bias is balanced out 
by assigning each conformation a Rosenbluth weight $W_n^{\rm R}\sim p_n^{-1}$.
Chain growth methods with population control such as PERM 
(Pruned-Enriched Rosenbluth Method)~\cite{grassberger1,grassberger2} 
and its recent modifications nPERM$^{\rm ss}_{\rm is}$~\cite{hsu1} 
improve this procedure considerably by utilising the counterbalance between Rosenbluth weight 
and Boltzmann probability.
The weight factor $W_n^{\rm R}$ is therefore replaced by 
\begin{eqnarray}
\label{PERMweight}
&&W_n^{\rm PERM}=\prod\limits_{l=2}^n m_l e^{-(E_l-E_{l-1})/k_BT},\\
&& 2\le n\le N \quad (E_1=0, \quad W_1^{\rm PERM}=1), \nonumber
\end{eqnarray} 
where $E_l$ is the energy of the partial chain 
${\bf X}_l=({\bf x}_1,\ldots ,{\bf x}_l)$ created with Rosenbluth chain growth and $T$
is the temperature. 

To explain the main ideas, we shall confine ourselves for the moment 
to the original PERM formulation~\cite{grassberger1}, where the sample of chains of length $n$ is enriched
by making identical copies once
$W_n^{\rm PERM}$ is bigger than a certain threshold value 
$W_n^>$.
In this case, the weight $W_n^{\rm PERM}$ is divided among the clones. 
For $W_n^{\rm PERM}$ being
smaller than a lower bound $W_n^<$, the chain is pruned with probability $1/2$ and the weight of 
a surviving chain is doubled.
The partition sum is proportional to the sum of weight 
factors (\ref{PERMweight}) for the conformations ${\bf X}_{n,t}$ of length $n$ 
sampled at ``time'' $t$,
\begin{equation}
\label{PERMpartsum}
Z_n\sim\sum\limits_{t} W_n^{\rm PERM}({\bf X}_{n,t}).
\end{equation}
The PERM algorithms are very successful as ground-state searchers and the canonical distribution 
at a given temperature $T$ is well reproduced over some orders of magnitude, but states that are 
highly suppressed at this temperature are not hit in a reasonable time. Standard reweighting 
techniques are applicable only in a small region around $T$. Thus, recording temperature-dependent
quantities such as the specific heat requires simulations at different temperatures.

As the partition sum of a polymer or a heteropolymer with fixed sequence
can be expressed in terms of the density (or degeneracy) of states $g(E)$,
$Z = \sum_{\{{\bf x}\}} e^{-\beta E(\{{\bf x}\})} = \sum_i g(E_i)e^{-\beta E_i}$
($\beta\equiv 1/k_BT$), all energetic quantities such as
the mean energy $\langle E\rangle(T)=-(\partial/\partial\beta) \ln Z$, the specific
heat $C_V(T)=\left(\langle E^2\rangle-\langle E \rangle^2\right)/k_BT^2$,
Gibbs free energy $F(T)=-k_BT\ln Z$, and entropy $S(T)=\left(\langle E \rangle-F\right)/T$
can directly be calculated if the density of states is known. These quantities are of 
particular interest, since they are indicators of temperature-dependent
conformational transitions. 

Our method allows within one simulation the direct sampling
of the density of states $g(E)$ over the entire range of the energy space 
with probabilities ranging over
many orders of magnitude, due to 
combining the advantages of energetic flat histogram reweighting at infinite
temperature and chain growth. 
The flat distribution can then be reweighted to any desired temperature.
Rare, i.e.\ low-lying energy states are also hit and 
therefore the low-temperature behaviour of the polymer can be reproduced well, in particular 
the low-temperature transition between compact globules and ground states of 
lattice proteins with low ground-state degeneracy. Using the HP model, we applied the method to 
lattice proteins with more than 40 monomers and different ground-state degeneracies
and found for examples with low ground-state degeneracy pronounced low-temperature peaks 
in the specific heat indicating ground-state -- globule transitions. 
Since our method is completely general, it is also applicable to other polymer models.  

In order to achieve a flat distribution of energetic states using chain growth, 
we introduce into the partition sum (\ref{PERMpartsum}) 
an additional weight $W_n^{\rm flat}(E_n({\bf X}_n))$ 
that depends on the energy $E_n$ of a given conformation ${\bf X}_n=({\bf x}_1,\ldots,{\bf x}_n)$:
\begin{eqnarray}
\label{flatpartsum}
&&Z_n\sim\nonumber \sum\limits_{t} W_n^{\rm PERM}({\bf X}_{n,t})\nonumber \\
&&\hspace{15mm} \times\, W_n^{\rm flat}(E_n({\bf X}_{n,t})) 
\left[W_n^{\rm flat}(E_n({\bf X}_{n,t}))\right]^{-1}.
\end{eqnarray}   
Since the histograms at all intermediate stages of the chain growth process are required to be flat, the 
new reweighting factor is rewritten in product form and we have
\begin{eqnarray}
\label{flatpartsumb}
&&Z_n\sim\sum\limits_{t} \left[W_n^{\rm flat}(E_n)\right]^{-1}\nonumber\\
&&\hspace{17mm}\times\, \prod\limits_{l=2}^n m_l
e^{-(E_l-E_{l-1})/k_BT} \frac{W_l^{\rm flat}(E_l)}{W_{l-1}^{\rm flat}(E_{l-1})}
\end{eqnarray}
with $W_1^{\rm flat} = 1$. The PERM weight factors (\ref{PERMweight}) lead to a canonical distribution $P^{{\rm can},T}_n(E_n)$ 
which shall be deformed to a constant distribution $P^{{\rm flat},T}_n(E_n)$ over the entire energy space. This 
requires the weights $W_n^{\rm flat}$ to be proportional to the inverse of the canonical distribution,
$W_n^{\rm flat}\sim 1/P^{{\rm can},T}_n(E_n)$, a condition that can obviously only be satisfied
iteratively~\cite{muca}. As we are mainly interested in the density of states,
which is proportional to the canonical probability distribution at $\beta= 1/k_BT=0$, 
$g_n(E_n)\sim P^{{\rm can},\infty}_n(E_n)$, we will effectively perform the simulation at infinite temperature. 
Consequently, $W_n^{\rm flat}\sim 1/g_n(E_n)$ and $Z_n\sim\sum_{t} g_n(E_n({\bf X}_{n,t})) W_n({\bf X}_{n,t})$.
Here we have introduced the combined weight
\begin{equation}
\label{combweight}
W_n({\bf X}_n) = \prod\limits_{l=2}^n m_l \frac{g_l^{-1}(E_l)}{g_{l-1}^{-1}(E_{l-1})}, \quad W_1= g_1=1,
\end{equation}
which can also be written recursively, $W_n=W_{n-1} m_n g_n^{-1}(E_n)/ g_{n-1}^{-1}(E_{n-1})$.
The canonical mean value of any quantity $O_n({\bf X}_n)$ is then calculated as follows:
\begin{eqnarray}
\label{meanvalue}
\langle O_n\rangle_{\rm can} &=& \frac{\sum\limits_{t} O_n({\bf X}_{n,t}) g_n(E_n({\bf X}_{n,t})) 
W_n({\bf X}_{n,t})}
{\sum\limits_{t} g_n(E_n({\bf X}_{n,t})) W_n({\bf X}_{n,t})}\nonumber\\
&=&
\frac{\langle O_n g_n(E_n)\rangle_{\rm flat}}{\langle g_n(E_n)\rangle_{\rm flat}},
\end{eqnarray}
where we have used mean values according to the flat distribution 
$\langle\ldots \rangle_{\rm flat}=\sum_{t}\ldots W_n({\bf X}_{n,t})/Z_n^{\rm flat}$ with
the partition sum of the flat distribution $Z_n^{\rm flat}=\sum_{t} W_n({\bf X}_{n,t})$.
The most important technical part of the algorithm is the determination of the weights $W_n^{\rm flat}$,
since they are directly connected with the desired densities of states $g_n(E)$. As the weights are completely 
unknown in the beginning, we evaluate them iteratively, starting from unity, $W_n^{{\rm flat},(0)}(E)=1$ 
($2\le n\le N$) for all
values of $E$. This means that the zeroth iteration is a pure chain growth run without reweighting. 

Each time, a chain of length $n$ with energy $E$ is created, the corresponding
histogram value $h_n(E)$ is increased by the weight $W_n$ of the chain. If a certain number of chains with total
length $N$ was produced, the iteration is finished and the new weights $W_n^{{\rm flat},(i)}(E)$ are determined
by calculating $W_n^{{\rm flat},(i)}(E)=W_n^{{\rm flat},(i-1)}(E)/h_n(E)$, $2\le n\le N$. Before
the next iteration starts, the histogram is reset, $h_n(E)=0$.
The correct weights are found, when $h_n(E)$ is ``flat'', i.e., it has approximately the same value for all energies.
In our actual implementation we employed a suitably adapted multicanonical 
variant of nPERMis ({\em n}\/ew PERM with {\em i}\/mportance {\em s}\/ampling)~\cite{hsu1}. 

Before we present results obtained with the new algorithm, we discuss general properties of heteropolymers
exemplified for 14mers which can still be analysed by exact enumeration.
Among all $2^{14}$ sequences for 14mers there is only one designing sequence, i.e.\ a
sequence to which a unique ground state belongs (up to a reflection symmetry).
\begin{figure}[t]
\centerline{
\epsfxsize=8.5cm \epsfbox{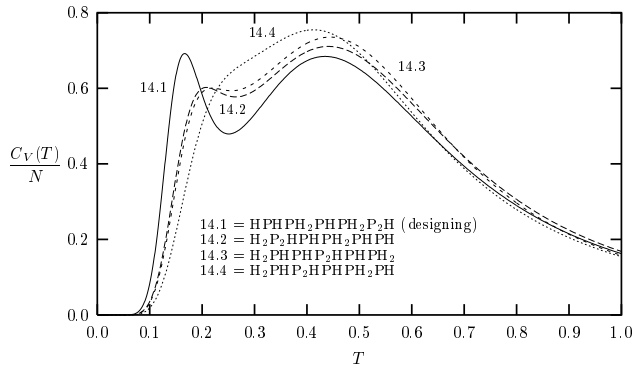}
}
\caption{\label{14mers}
Specific heat for four 14mers with different sequences. 
Only the 14mer with the native ground state
shows a pronounced low-temperature peak indicating a transition between ground-states and globules.
}
\end{figure}
We compared thermodynamic properties of four 14mers with different sequences but same hydrophobicity ($n_H=8$)
and identical lowest energy ($E_{\min} = -8$). 
Figure~\ref{14mers} shows the specific heat for the different 14mers. A pronounced 
low-temperature peak that
indicates the transition between ground states and compact globule states is only observed
for the 14mer with the native ground state, 14.1.
These results show qualitatively how the conformational transitions depend on the ground-state 
degeneracy $g_0$ of the polymer. For the sequences 14.2 and 14.3 it is twice that of the designing
sequence, and 14.4 is even four times higher degenerate. 

\begin{figure}[t]
\centerline{
\epsfxsize=8.5cm \epsfbox{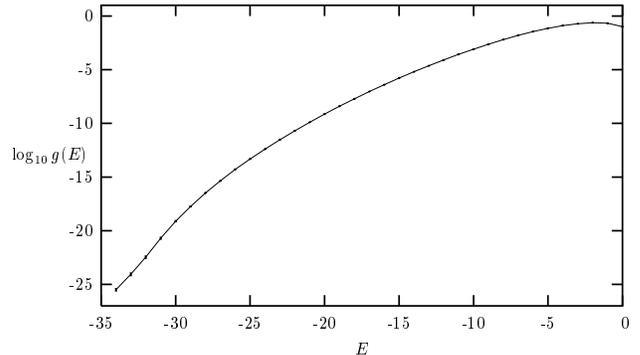}
}
\caption{\label{42mer_gE}
Density of states of the 42mer, normalised to unity, i.e.\ $\sum_i g(E_i)=1$.} 
\end{figure}

The first example to which we applied our multicanonical chain growth algorithm is a 42mer
with the sequence PH$_2$PHP\-H$_2$\-P\-H\-P\-H\-P$_2$\-H$_3$\-P\-H\-P\-H$_2$\-P\-H\-P\-H$_3\-$P$_2$\-H\-P\-H\-P\-H$_2$\-P\-HPH$_2$P whose 
ground-state properties
have similarities with the parallel $\beta$-helix of {\em pectate lyase C}~\cite{yoder}.
The lattice model with only four-fold ground-state degeneracy has a ground-state
energy of $E_{\min}=-34$. In order to investigate the low-temperature behaviour of this
system it is necessary that the algorithm correctly samples the low-energy states
and that it also hits the ground states. 
The measured density of states ranges over about 25 orders of magnitude and
covers the entire energy space $[-34,0]$, as shown in Fig.~\ref{42mer_gE}. 
\begin{figure}[b]
\centerline{
\epsfxsize=8.5cm \epsfbox{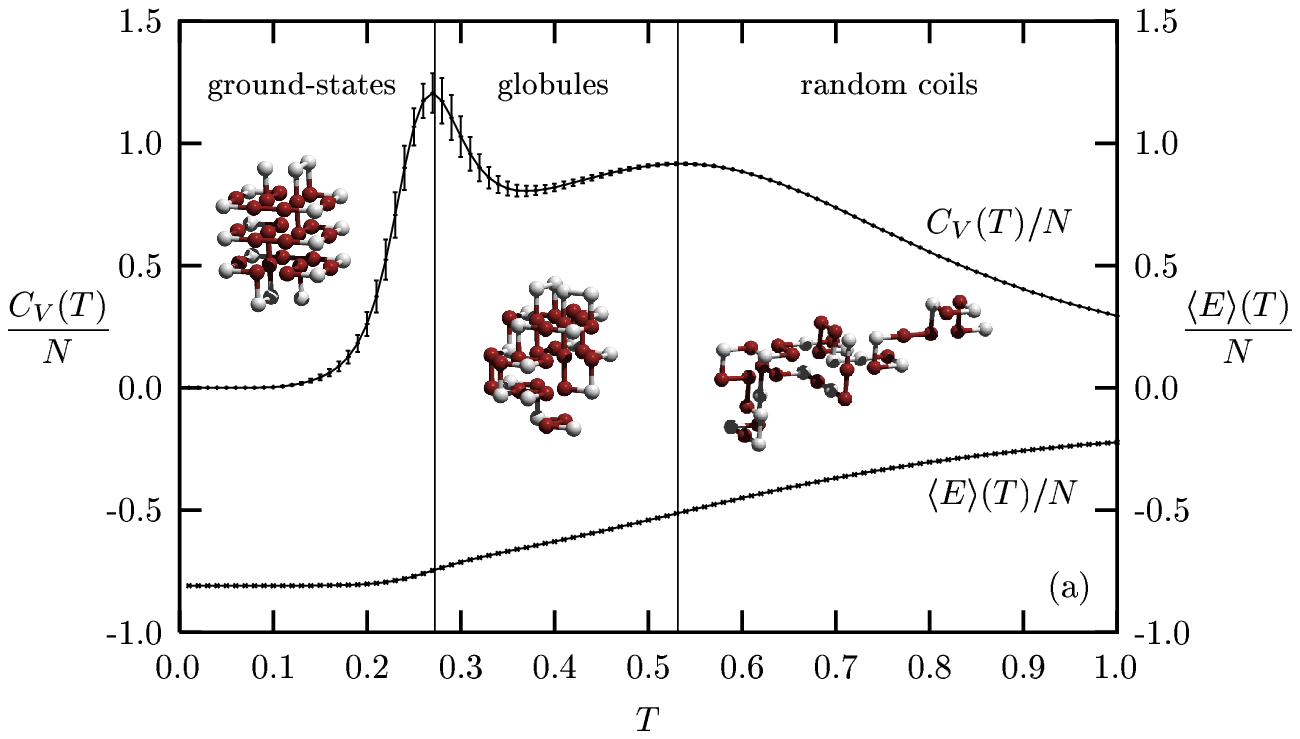}
}
\centerline{
\epsfxsize=8.5cm \epsfbox{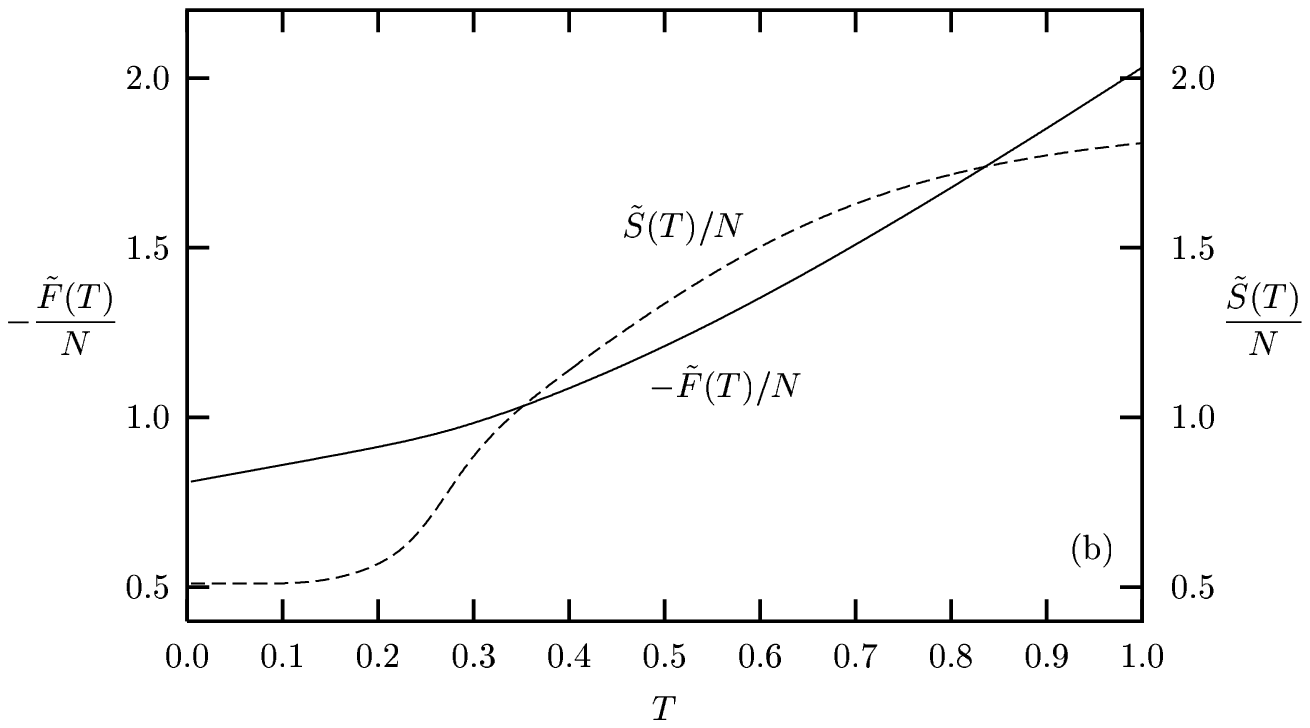}
}
\caption{\label{42mer_CEFS}
(a) Specific heat and mean energy as functions of the temperature for the 42mer. 
The ground-state -- globule
and the globule -- random coil transition occur at $T_0\approx 0.27$ and $T_1\approx 0.53$,
respectively. (b) Reduced free energy $\tilde{F}(T)=F(T)+TS_0$ and entropy $\tilde{S}(T)=S(T)-S_0$.}
\end{figure}
Figure~\ref{42mer_CEFS}(a) shows the results for the heat capacity and the mean energy. 
Writing out the raw energies and weights from the simulation, we analysed the data
and calculated the statistical error by means of the jackknife blocking method.
Due to the
low degeneracy of the ground states, the transition between native states and globule
states is very pronounced and occurs at a temperature $T_0\approx 0.27$. The 
globule -- random coil transition at $T_1\approx 0.53$, on the
other hand, is rather weak. This confirms the results of Ref.~\cite{iba1}. 
From the density of states, we also obtain the
Gibbs free energy $F(T)$ and the entropy $S(T)$. 
Since the ground-state degeneracy $g_0$ is not accessible
by stochastic search algorithms in general, we have plotted in Fig.~\ref{42mer_CEFS}(b) the reduced
free energy $\tilde{F}(T)=F(T)+TS_0$ and the entropy $\tilde{S}(T)=S(T)-S_0$,
up to the constant $S_0=\ln g_0$, where $g_0$ is the degeneracy of the ground state.

\begin{figure}
\centerline{
\epsfxsize=8.5cm \epsfbox{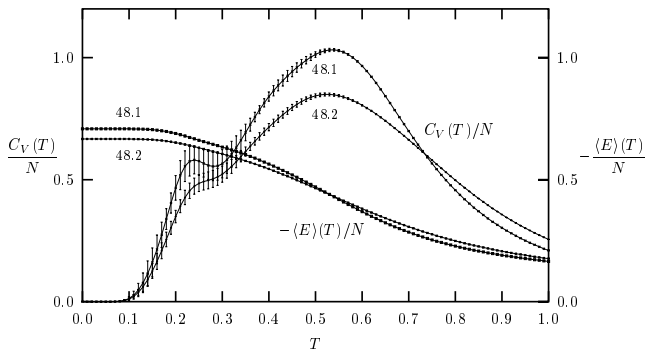}
}
\caption{\label{48mers_CE}
Specific heat and mean energy for the two 48mers with different ground-state
properties. The low-temperature conformational transition is more pronounced
for the example 48.1 with lower ground-state degeneracy.}
\end{figure}

Finally, similar to the consideration of the 14mers, 
we compare two
48mers with different ground-state properties. The first one,
which we denote by $48.1$ has the sequence 
PH\-P\-H$_2$\-P\-H$_6$\-P$_2$\-H\-P\-H\-P$_2$\-H\-P\-H$_2$\-P\-H\-P\-H\-P$_3$\-H\-P$_2$\-H$_2$\-P$_2$\-H$_2$\-P$_2$\-H\-P\-H\-P$_2$\-HP
and its ground state with the energy $-34$ is $5000$-fold degenerate.
The ground state of the other 48mer ($48.2$) with
the sequence
HP\-H$_2$\-P$_2$\-H$_4$\-P\-H$_3$\-P$_2$\-H$_2$\-P$_2$\-H\-P\-H$_3$\-P\-H\-P\-H$_2$\-P$_2$\-H$_2$\-P$_3$\-H\-P$_8$H$_2$
has the much higher degeneracy 
of $1.5\times 10^6$ and possesses the energy $-32$~\cite{dill3}.
As is demonstrated in Fig.~\ref{48mers_CE}, we also observe for these longer chains that 
the conformational transition between the lowest-energy states and
the globules is stronger the lower the ground-state degeneracy 
is.
In conclusion, we have developed a multicanonical chain growth algorithm that 
allows the simulation of the thermodynamic properties of
polymers and heteropolymers. It is based on energetic
flat histogram sampling of the density of states in combination with 
PERM chain growth. We applied this algorithm to heteropolymers
with more than 40 monomers and obtained accurate  
densities of states over about 25 orders of magnitude that
cover the entire energy range, thus yielding
very good results for all derived energetic quantities such
as mean energy, specific heat, free energy, and entropy.
In particular, this enabled us to determine the low-temperature
behaviour of the systems with high precision and to observe pronounced
low-temperature peaks of the specific heat for lattice
proteins with low ground-state degeneracy indicating the
ground-state -- globule transition. \\[5mm]
We thank Peter Grassberger and Hsiao-Ping Hsu for helpful discussions on the 
new PERM algorithms.
This work is partially supported by the German-Israel-Foundation (GIF) under
contract No.\ I-653-181.14/1999. 
\end{document}